\documentclass[reprint,
onecolumn,
 amsmath,amssymb,
 aps,
 pra,
]{revtex4-2}

\usepackage{graphicx}
\usepackage{dcolumn}
\usepackage{bm}
\usepackage{xcolor}
\usepackage{hyperref}

\newtheorem{teo}{Theorem} 
\newtheorem{cor}{Corollary} 
\newtheorem{prop}{Proposition} 

\newcommand{\Cov}{\mathrm{Cov}}

\newcommand{\Real}{\mathrm{Re}}
\newcommand{\Imag}{\mathrm{Im}}

\newcommand{\eye}{\mathbb{I}}
\newcommand{\R}{\mathbb{R}}
\newcommand{\C}{\mathbb{C}}
\newcommand{\Abs}{\mathrm{Abs}}
\newcommand{\ket}[1]{\ensuremath{|\,{#1}\,\rangle}}
\newcommand{\bra}[1]{\ensuremath{\langle\,{#1}\,|}}

\newcommand{\ev}[1]{\ensuremath{\left\langle\,{#1}\,\right\rangle}}

\newcommand{\dyad}[2]{\ensuremath{|\,{#1}\,\rangle\langle\,{#2}\,|}}

\newcommand{\rmi}{\mathrm{i}}
\newcommand{\Tr}{\mathrm{Tr}}

\begin{document}

\preprint{APS/123-QED}

\title{Complex Field Formulation of Quantum Estimation Theory}

\author{M. Muñoz}
\email{marioamunoz@udec.cl}
\affiliation{School of Applied Mathematics, Fundação Getulio Vargas, Rio de Janeiro, Brazil}

\author{L. Pereira}
\affiliation{Instituto de Física Fundamental IFF-CSIC, Calle Serrana 133b, Madrid 28006, Spain, and ICFO - Institut de Ciencies Fotoniques, The Barcelona Institute of Science and Technology, 08860 Castelldefels, Barcelona, Spain}

\author{C. Vargas}
\affiliation{Instituto Milenio de Investigación en Óptica y Departamento de Física, Facultad de Ciencias Físicas y Matemáticas, Universidad de Concepción, Casilla 160-C, Concepción, Chile}

\author{S. Niklitschek}
\affiliation{Departamento de Estadística, Facultad de Ciencias Físicas y Matemáticas, Universidad de Concepción, Casilla 160-C, Concepción, Chile}

\author{A. Delgado}
\email{aldelgado@udec.cl}
\affiliation{Instituto Milenio de Investigación en Óptica y Departamento de Física, Facultad de Ciencias Físicas y Matemáticas, Universidad de Concepción, Casilla 160-C, Concepción, Chile}

\date{\today}

\begin{abstract}
We present a complex field formulation of the quantum estimation theory that works natively with complex statistics on the dependence of complex parameters. This formulation states new complex versions of the main quantities and results of the estimation theory depending on complex parameters, such as Fisher information matrices and Cram\'er-Rao bounds. This can be useful in contexts where the quantum states are described through complex parameters, such as coherent states or squeezed states. We show an example of an application of our theory in quantum communication with coherent states.
\end{abstract}


\maketitle

\section{Introduction}

The accurate determination of quantities plays a key role in the development of physical theories and their applications. For instance, high-precision measurements of the gyromagnetic factors of the electron and muon, and their comparison with theoretical predictions, are important low-energy tests of relativistic quantum field theory of electrodynamics \cite{Aoyama2019,Keshavarzi2022,Fan2023}. In this context, estimation theory has become an ubiquitous tool nowadays. This theory, which deals with the estimation of the parameters values based on experimentally acquired data that exhibit randomness, leads to the celebrated Cram\'er-Rao bound for the best precision achievable \cite{Rao1992,cramer1946mathematical,Fisher1925}. A classic example of this bound is the standard or shot-noise limit for the interferometric determination of a phase. 

Over the last two decades, advances in quantum information theory and the foundations of quantum mechanics \cite{nielsen2010quantum} have led to considering quantum states and measurements as resources that can further improve the accuracy of estimation processes. This has motivated the formulation of a quantum estimation theory \cite{Helstrom1969,Belavkin1976,Paris2008} that finds application when a relevant quantity is encoded in a quantum state and must be estimated using data obtained by quantum measurements. This theory plays a relevant role in quantum metrology \cite{Giovannetti2011,Toth2014,Szczykulska2016,Giovannetti2006}, quantum sensing \cite{Degen2017}, quantum natural gradient \cite{Stokes2020,Gacon2021,Meyer2021, Gidi2023}, and quantum tomography \cite{Mahler2013, Hou2016, Li2016, Pereira2018, Struchalin2018, 2009.04791, Zambrano2020, Zambrano2020_3B}, and has already been experimentally implemented in optical interferometry \cite{Caves1981,Demkowicz-Dobrzanski2015,McCuller2021}, trapped ions \cite{Leibfried2004}, and condensed matter \cite{Wildermuth2006,Vengalattore2007}, among others. Quantum estimation theory enables us to exceed the standard limit and reach the Heisenberg limit, which corresponds to a quadratic increase in precision relative to the standard limit.

Estimation theory and its quantum extension are formulated over the field of real numbers. Despite the latter being successful \cite{Liu2020, Yuen1973, Fujiwara1999, Hayashi2008, Petz_2002, Suzuki2016}, this adaptation seems unnatural, since quantum mechanics is naturally formulated in the field of complex numbers. For instance, coherent and squeezed states are described by complex coefficients that are functions of complex variables, and quantum processes and measurements are described by the complex-valued Choi matrices \cite{Stricker2022}. Recently, the advantages of estimating complex quantities using their native representation have been studied in the case of radio interferometric gain calibration \cite{Smirnov2015}, and optimization of complex variables has been applied to signal analysis \cite{Adali2014,kay1993, Bickel2007, Gorman1990, Marzetta1993, Yau1992, vandenBos1994, Carvalho2000, Smith2005} and neural networks \cite{Hirose2012, Zhang2015}. Furthermore, complex numbers have been shown to play a fundamental role in the formulation of quantum mechanics \cite{2101.10873,Chen2022}, that is, a formulation of quantum mechanics on real numbers leads in certain scenarios to predictions different from those of the quantum mechanics formulated on complex numbers. Alternative formulations of estimation theory have been proposed \cite{Ollila2008, Jagannatham2004}, where the Fisher information matrix and the Cram\'er-Rao bound are adapted to consider complex statistics of complex parameters. This is implemented using Wirtinger complex calculus \cite{WirtingerZurFT,Kreutz2009} and a particular complex map \cite{Ollila2008}. The Wirtinger calculus is developed on the joint basis of complex variables and their conjugates, thereby enabling a complex differentiation theory analogous to that of their real counterparts. This approach has been used to estimate pure quantum states by optimization in the field of complex numbers \cite{Utreras2019, Zambrano2020}. The complex map connects the main results of the theory of quantum estimation formulated in real numbers with those formulated in complex numbers, such as the Fisher information matrix and the Cram\'er-Rao bound.

Motivated by the above, in this article we develop a complex-field formulation of quantum estimation theory based on Wirtinger calculus, working with complex statistics that depend on complex parameters. This formulation provides a natural framework for studying estimation problems in which the quantum states are native complex functions of complex parameters. We define complex-field versions of the logarithmic derivatives, quantum Fisher information matrices, the Cram\'er-Rao bound, and bounds on the weighted mean-square error, both symmetric and right. We also specialize all the above results to pure states. We apply our theory to an example in quantum communications, that is, the optimal estimation of a complex parameter encoded in a coherent state of the electromagnetic field \cite{Arnhem2019}.  

The paper is organized as follows: In section II, we introduce some essential preliminary results such as the Wirtinger Calculus, the estimation theory for complex statistics dependent on complex parameters, and the quantum estimation theory. In section III, we develop the complex field formulation of the quantum estimation theory. In section IV, we apply our formulation to the problem of optimal quantum communications using a coherent state. In section V, we conclude and summarize.

\section{Preliminary results}

In this section, we briefly review the main results of estimation theory. First, we introduce the estimation theory based on real parameters, which studies complex statistics through a real transformation of themselves and their parameters. We present recent results on the complex statistics estimation problem for complex parameters \cite{Kreutz2009}. In the case of estimating parameters from quantum states, the quantum properties can be exploited to formulate a quantum version of the estimation theory. We also summarize the basic elements of the quantum estimation theory.

\subsection{Estimation theory for complex parameters}\label{section1}

Estimation theory, in a simplified way, deals with the problem of estimating the value of a parameter $\theta$ based on empirical data measured for a random variable $X$ \cite{cramer99}. Specifically, it is considered a random variable $X$ with values in a measurable space $\left( \Omega, \mathcal{F}\right)$, called a sample space, defined by a set $\Omega$ and a $\sigma$-algebra $\mathcal{F}$ on $\Omega$. The   variable $X$ is characterized by its probability distribution function $\mathrm{P}_X$, which is known to belong to a class of distributions $\textbf{P}=\{\mathrm{P}_\theta: \theta \in \Theta \subset \mathbb{K}^n\}$, where $\mathbb{K}\in \left\{\mathbb{R},\mathbb{C}\right\}$. So, there is a parameter $\theta= \left[\theta_1,\theta_2,\dots,\theta_k \right] \in \Theta$ such that $\mathrm{P}_X = \mathrm{P}_\theta $. Thus, the problem of parameter estimation consists in approximating some quantity $\Phi(\theta)$ by the mean of some statistic $t:\mathcal{X}\to\Phi(\Theta)$ called the estimator, where $\mathcal{X}=[x_1,x_2\dots x_N]$ is the total data from the observations of $X$.

Consider the problem of estimating a complex parameter $\theta=\alpha+i\beta\in\C^k$, with $\alpha$ and $\beta$ its real and imaginary parts, using a complex random variable $X$. This is usually stated in the field of real numbers. We define the real representation $\bar z$ of a complex quantity $z=x+iy\in\C^k$ as the stacking of its real and imaginary parts, that is,
\begin{equation}
\bar{z}=\left[\begin{array}{c}x\\ y \end{array}\right]\in \R^{2k}.
\end{equation}
Let $f(\omega|\theta)$ be the probability density function for obtaining the random variable $X$ in the value $\omega\in\Omega$ given the parameter $\theta$. Notice that this probability density function does not depend on the representation of the parameter $\theta$, so that $f(\omega|\theta)$ and $f(\omega|\bar\theta)$ are equivalent. To obtain an estimator of $\theta$ we consider a complex statistic $t$, whose real representation $\bar t$ has a covariance matrix 
\begin{equation}
\Cov_{\bar\theta}\big(\bar{t}\big)= \sum_{\omega\in\Omega} f(\omega|\bar\theta) \big(\bar{t}_{\bar\theta}(\omega)-\bar{\mu}_{\bar\theta}\big)\big(\bar{t}_{\bar\theta}(\omega)-\bar{\mu}_{\bar\theta}\big)^\top ,
\end{equation}
where $t_{\bar\theta}(\omega)$ is the estimator of $\bar\theta$ given the outcome $\omega$ and $\mu_\theta$ is the expected value of $\bar t$, 
\begin{equation}
    \bar{\mu}_{\bar\theta} = \sum_{\omega\in\Omega} f(\omega|\bar\theta)\bar{t}_{\bar\theta}(\omega) .
\end{equation}
The Fisher information matrix (FIM) of $\bar\theta$ is defined by
\begin{equation}\label{CFIM}
I_{\bar\theta}=\sum_{\omega\in\Omega} f(\omega|\bar\theta) \nabla_{\bar\theta}\ln f(\omega|\bar\theta)\thinspace\nabla_{\bar\theta}\ln f( \omega|\bar\theta)^\top.
\end{equation}
The Cram\'er-Rao inequality (CRI) for $\bar{t}$ states that
\begin{equation}\label{ClassicCRB}
\Cov_{\bar\theta}\left(\bar{t}\right)\geq D_{\bar\mu}(\bar\theta)I_{\bar\theta}^{-1}D_{\bar\mu}(\bar\theta)^\top,
\end{equation}
where $D_{\bar\mu}$ is the Jacobian matrix of $\bar{\mu}_{\bar\theta}$. The term on the right side of (\ref{ClassicCRB}) is the Cram\'er-Rao bound (CRB) and represents the smallest uncertainty with which the statistic $\bar t$ can be measured.

The above analysis can be replicated working in its native complex field, as was shown in \cite{Ollila2008}. Consider the continuous map $\ev{\cdot}_\C:\mathbb{C}^{2d\times2k}\rightarrow\mathbb{C}^{2d\times2k}$, defined for all $G \in\mathbb{C}^{2d\times2k}$ by
\begin{equation}\label{Cmap}
\ev{G}_\C=2M_{2d}^{-1} G M_{2k},
\end{equation}
where $M_{2l}$ is a complex matrix of size $2l\times 2l$ given by
\begin{equation}
M_{2l}:=\frac{1}{2}\left[\begin{array}{cc}\eye_l& \eye_l\\-\rm{i}\eye_l&\rm{i}\eye_l \end{array}\right],
\end{equation}
for any $l\in \mathbb{N}$, and $\eye_l$ is the identity matrix $l\times l$. Note that $M_{2l}$ is invertible, its inverse is given by $M_{2l}^{-1}=2M_{2l}^\dagger$. The action of the map onto a block matrix
\begin{equation}
    G=\left[\begin{array}{cc}G_{11}&G_{12}\\G_{21}&G_{22}\end{array}\right]
\end{equation}
where $G_{jk}\in\C^{d\times k}$ leads to
\begin{equation}\label{MBM}
 \ev{G}_\C=\left[\begin{array}{cc}G_{11}+\rmi G_{21}-\rmi(G_{12}+\rmi G_{22})& G_{11}+\rmi G_{21}+\rmi(G_{12}+\rmi G_{22})\\
G_{11}-\rmi G_{21}-\rmi(G_{12}-\rmi G_{22})& G_{11}-\rmi G_{21}+\rmi(G_{12}-iG_{22})\end{array}\right].
\end{equation}
Other properties of the map $\ev{\cdot}_\C$, which will be very useful later, are presented in the following proposition \cite{Ollila2008}:
\begin{prop}\label{Prop of the map}
Let $G,H \in \mathbb{C}^{2d\times 2k}$, $\sigma_{2k}$ a rotation matrix given by 
\begin{equation}
\sigma_{2k}=\left[\begin{array}{cc} 0 & \mathbb{I}_{2k}\\ \mathbb{I}_{2k}&0\end{array}\right],
\end{equation}
and $f$  a function defined on the spectrum of a matrix, that is,
\begin{equation}
f(G)=U^{-1} f(\Lambda) U,\qquad \forall \thinspace G \in \mathbb{C}^{2d\times 2k},
\end{equation}
where $G=U^{-1}\Lambda U$, with $\Lambda$ the diagonal matrix with the eigenvalues of $G$. Then
\begin{enumerate}
\item \begin{equation}\label{property4}
 \ev{G+\lambda H}_\C=\ev{G}_\C+\lambda\ev{H}_\C, \  \forall \thinspace \lambda \in \mathbb{C}.
\end{equation}
\item \begin{equation}\label{property1} 
 G=G^\dagger \Longleftrightarrow  \ev{G}_\C=\ev{G}_\C^\dagger.
\end{equation}
\item \begin{equation}\label{property2}
 G \ invertible \Longleftrightarrow \ev{G}_\C  invertible.\ Moreover \ \ev{G}_\C^{-1}=\frac{1}{4}\ev{G^{-1}}_\C.
\end{equation}
\item \begin{equation}\label{property6} 
 G\geq 0 \Longleftrightarrow \ev{G}_\C\geq0,\ for\ each\ symmetric\ matrix\ G.
\end{equation}
\item \begin{equation}\label{property3} 
 \ev{G^\dagger}_\C=\ev{G}_\C^\dagger.
\end{equation}
\item \begin{equation}\label{property5}
\ev{G_1G_2}_\C=\frac{1}{2}\ev{G_1}_\C\ev{G_2}_\C.
\end{equation}
\item \begin{equation}\label{property7}
 \Tr(G)=\frac{1}{2}\Tr(\ev{G}_\C).
\end{equation}
\item \begin{equation}\label{property9}
 \ev{f(G)}_\C = 2 f\left(\frac{1}{2} \ev{G}_\C \right).
\end{equation}
\item \begin{equation}
 \ev{G^*}_{\C} = \sigma_{2d} \ev{G}^*_{\C}\sigma_{2k}.
\end{equation}
\end{enumerate}
\end{prop}

The concept of the derivation of a function of complex variables given by complex calculus \cite{martin1968, Scott1985} is limited to holomorphic functions, that is, functions that are infinitely differentiable in a neighborhood of the point. This is a strong restriction that leaves many interesting complex functions without derivatives, for example $f(z)=|z|^2$. However, there is a generalization of the derivation of complex functions through the Wirtinger operators or Wirtinger derivatives \cite{Kreutz2009}, which are defined by
\begin{equation}\label{Wirtinger derivatives}
\partial_z=\frac{1}{2}\big(\partial_x-\rmi\partial_y\big),\qquad \partial_{z^*}=\frac{1}{2}\big(\partial_x+\rmi\partial_y\big),
\end{equation}
where $z=x+\rmi y\in \C$, and $z^*$ denote its complex conjugate. These operators can be applied to any function $f(z)=f(z,z^*)=f(x,y)$ whose real and imaginary parts have partial real derivatives. An important property of these operators lies in their behavior, which emulates the partial derivative of a real-valued function of a real variable \cite{Kreutz2009}. 

We define the conjugated extension $\hat{z}$ of a complex quantity $z\in\C^k$ as the stacking of $z$ and its conjugate $z^*$, that is,
\begin{equation}
\hat{z}=\left[\begin{array}{c}
z\\z^*\end{array}\right]\in\C^{2k}.
\end{equation}
Notice that $\bar{z}=M_{2d}\hat{z}$. Then, for the case of a multivariable complex function $F:\mathbb{C}^n\longrightarrow \mathbb{C}^m$, its Jacobian matrix is defined by 
\begin{equation}\label{complex jacobian}
\mathcal{D}_{\hat F}(z_0)  := \mathcal{D}_{\hat z}\hat F(z_0) = \left[\begin{array}{cc}
            \mathcal{D}_z F (z_0) &\mathcal{D}_{z^*} F(z_0) \\
            \big(\mathcal{D}_{z^*}F(z_0)\big)^*&\big(\mathcal{D}_{z}F(z_0)\big)^*
            \end{array}\right] \in \mathbb{C}^{2m\times 2n},
\end{equation}
for each $z_0\in \mathbb{C}^n$, and
\begin{equation}
\left[\mathcal{D}_z F (z_0)\right]_{ij}=\partial_{z_j} F^i(z_0),\qquad \left[\mathcal{D}_{z^*}F(z_0)\right]_{ij}=\partial_{z^*_j} F^i(z_0),
\end{equation}
for $i=1,\dots,n$ and $j=1,\dots,m$. Note that $D_{\hat F}$ involves the derivatives of $F$ and also of $F^*$, because $\big(\mathcal{D}_{z^*}F(z_0)\big)^*=\mathcal{D}_{z}F^*(z_0)$ and $\left(\mathcal{D}_{z}F(z_0)\right)^*=\mathcal{D}_{z^*}F^*(z_0)$. The Wirtinger Jacobian is related to the real Jacobian with complex map $\ev{\cdot}_\C$ by
\begin{equation}\label{Jacobian equiv}
\mathcal{D}_{\hat F}(z)=\frac{1}{2}\ev{D_{\bar{F}}(z)}_{\mathbb{C}}.
\end{equation}

Using the conjugate extension and the Wirtinger calculus, we can construct the complex version of the estimation theory. The covariance matrix of the conjugate extension $\hat{t}$ of the statistic $t$ is defined by
\begin{equation}
\Cov_{\hat{\theta}}(\hat{t}) = \sum_{\omega\in\Omega} f(\omega|\hat\theta) \big(\hat{t}_{\hat\theta}(\omega)-\hat{\mu}_{\hat\theta}\big)\big(\hat{t}_{\hat\theta}(\omega)-\hat{\mu}_{\hat\theta}\big)^\dagger,
\end{equation}
with $\hat{\mu}_{\hat\theta} = \sum_{\omega\in\Omega} f(\omega|\hat\theta) \hat{t}_{\hat\theta}(\omega)$. The FIM for conjugate extension $\hat\theta$ is defined by
\begin{equation}\label{CCFIM}
\mathcal{I}_{\hat\theta}=\left[\begin{array}{cc}\mathtt{I}_\theta&\mathtt{P}_\theta\\\mathtt{P}_\theta^*&\mathtt{I}_\theta^*
\end{array}\right],
\end{equation}
where 
\begin{equation}
\mathtt{I}_\theta=\sum_{\omega\in\Omega}\nabla_{\theta^*}\ln f(\omega|\theta)\thinspace\nabla_{\theta^*}\ln f(\omega|\theta) ^\dagger  ,\quad
\mathtt{P}_\theta=\sum_{\omega\in\Omega}\nabla_{\theta^*}\ln f(\omega|\theta)\thinspace\nabla_{\theta^*}\ln f(\omega|\theta) ^\top ,
\end{equation}
are the FIM of $\theta$ and the pseudo-FIM (pFIM) of $\theta$, respectively, and  $\nabla_{\theta^*}=\left[\partial_{\theta^*_1},\partial_{\theta^*_2},\ldots,\partial_{\theta^*_n}\right]^\top$. Employing the complex map $\ev{\cdot}_\C$, we see that the covariance matrix and the FIM can be related to their real counterparts by
\begin{equation}\label{COV equiv}
\Cov_{\hat\theta}(\hat{t})=\ev{\Cov_{\bar\theta}\big(\bar{t}\big)}_\C,
\end{equation}
\begin{equation}\label{RCCFIM}
\mathcal{I}_{\hat\theta}=\frac{1}{4}\ev{I_{\bar{\theta}}}_{\mathbb{C}}.
\end{equation}
Applying the map $\ev{\cdot}_\C$ to \eqref{ClassicCRB} and using \eqref{COV equiv}, \eqref{Jacobian equiv} and \eqref{RCCFIM} we obtain the CRI to the complex statistic $t$, working directly on the complex field, that is, 
\begin{equation}\label{CClassicCRB}
\Cov_{\hat\theta}(\hat{t})\geq\mathcal{D}_{\hat{\mu}}\mathcal{I}_{\hat\theta}^{-1}\mathcal{D}_{\hat{\mu}}^{\dagger}.    
\end{equation}


\subsection{Quantum estimation theory for real parameters}\label{section_QET}

The aim of quantum estimation theory is to provide high-precision measurements to obtain valuable information about the system studied, exploiting its quantum properties \cite{Paris2008,Liu2020}.
Specifically, it is considered a finite-dimensional quantum system with Hilbert space $\mathcal{H} \cong \mathbb{C}^{d}$, the set of all bounded linear operators on $\mathcal{H}$ is denoted as $\mathcal{L}(\mathcal{H})\cong \mathbb{C}^{d\times d}$, and the space of observables (Hermitian matrices) as $\mathcal{L}_h(\mathcal{H})$. A quantum state is represented by a density operator $\rho\in \mathcal{L}_h(\mathcal{H})$ that satisfies $\rho \geq 0$ and $\mathrm{Tr}(\rho)=1$. The quantum state is parameterized as $\rho=\rho_\theta$ by a vector  $\theta = [\theta_1 ,\dots,\theta_n ]^\top \in \Theta \subseteq \mathbb{K}^n$ where $\mathbb{K}\in\left\{\mathbb{R},\mathbb{C}\right\}$, and the collection $\{\rho_\theta\}_{\theta\in\Theta}$ for all the values of $\theta$ is called a quantum statistical model. So, the goal of quantum estimation is to estimate some quantity $\Phi(\theta)$ by possibly measuring multiple copies of $\rho_\theta$.

Similarly to classical estimation theory, the standard approach to handling the estimation problem of a complex parameter with a complex statistic is to provide a real representation of both. Thus, the quantum state is parameterized by $\bar\theta = [\Real\,\theta,\Imag\,\bar{\theta}]^\top$, the quantity to be estimated is $\Phi(\bar\theta)$, and this value is estimated by $\bar{t} = [\Real\, t(\bar\theta),\Imag \, t(\bar\theta)]^\top$.

The probability density function is given by the Born rule $f(\omega|\bar\theta)=\Tr(\Pi_\omega\rho(\bar\theta))$, where $\boldsymbol{\Pi}=\{\Pi_\omega\}$
is the so-called positive operator-valued measurement (POVM), that is, a set of positive semi-definite operators that satisfy $\sum_{\omega\in\Omega} \Pi_\omega=\mathbb{I}$. The POVM becomes a new variable so that the covariance matrix $\Cov_{\bar\theta}(\boldsymbol{\Pi},\bar t)$ and the FIM $I_{\bar\theta}(\boldsymbol{\Pi})$ are also dependent on the POVM.

The symmetric logarithmic derivative (SLD) $L_{\bar\theta_i}^S$ of $\rho_{\bar\theta}$ with respect to $\bar\theta_i$ is implicitly defined by the equation \cite{Paris2008}
\begin{equation}\label{SLD}
\partial_{\bar\theta_i}\rho_{\bar\theta}=\frac{1}{2}\left(\rho_{\bar\theta} L^S_{\bar\theta_i}+L^S_{\bar\theta_i}\rho_{\bar\theta}\right),\qquad i=1,\ldots, 2k.
\end{equation}
Then, the SLD Quantum Fisher information matrix (SLD-QFIM) of the quantum state $\rho_{\bar\theta}$ for the real parameter $\bar \theta$ is given by
\begin{equation}\label{RSQFIM}
\Big[J^S_{\bar{\theta}}\Big]_{ij} =
\frac{1}{2}\Tr\Big(\rho_{\bar\theta}\big(L^S_{\bar\theta_i}L^S_{\bar\theta_j}+L^S_{\bar\theta_j}L^S_{\bar\theta_i}\big)\Big),\qquad i,j=1,\ldots, 2k.
\end{equation}

The right logarithmic derivative (RLD) $L_{\bar\theta_i}^R$ of $\rho_{\bar\theta}$ with respect to $\bar\theta_i$ is implicitly defined by the equation 
\begin{equation}\label{RQRLD}
\partial_{\bar\theta_i}\rho_{\bar\theta}=\rho_{\bar\theta} L^R_{\bar\theta_i}\qquad i=1,\ldots, 2k. 
\end{equation}
Then, the RLD Quantum Fisher information matrix (RLD-QFIM) of the quantum state $\rho$ for the real parameter $\bar \theta$ is given by \cite{Fujiwara1999,Hayashi2008}
\begin{equation}\label{RRQFIM}
\Big[J^R_{\bar{\theta}}\Big]_{ij} = 
\Tr\left(\rho_{\bar\theta} L^R_{\bar\theta_j}{L^R_{\bar\theta_i}}^\dagger\right),\qquad i,j=1,\ldots, 2k.
\end{equation}
Both QFIMs provide an upper bound for the FIM, that is $I_{\bar\theta}(\boldsymbol{\Pi}) \leq J^X_{\bar\theta}$, and define an extension of the CRI \eqref{ClassicCRB} to the case of quantum systems, called quantum Cramér-Rao inequality (QCRI), which establishes
\begin{equation}\label{QCRB}
\Cov_{\bar\theta}(\boldsymbol{\Pi},\bar{t}) \geq  D_{\bar\mu}(\bar\theta)I_{\bar\theta}(\boldsymbol{\Pi})^{-1}D_{\bar\mu}(\bar\theta)^\top \geq D_{\bar\mu}(\bar\theta)\left(J^X_{\bar\theta}\right)^{-1}D_{\bar\mu}(\bar\theta)^\top,
\end{equation}
for any $X \in \{S,R\}$, where $I_{\bar\theta}$ is the FIM of $\rho_{\bar\theta}$ for the real parameters $\bar \theta$ given by \eqref{CFIM}. The right side of the QCRI is known as the quantum Cram\'er-Rao bound (QCRB). Notice that while covariance matrix $\Cov_{\bar\theta}(\boldsymbol{\Pi},\bar{t})$ and FIM $I_{\bar\theta}(\boldsymbol{\Pi})$ depend on both state and measurement, QFIMs $J^X_{\bar\theta}$ only depend on the quantum state. This is because QCRB defines the minimum uncertainty over the set of all quantum measurements $\boldsymbol{\Pi}$ for estimating $\theta$ with quantum resources.

The QCRI \eqref{QCRB} is very useful for studying systems with few parameters. However, for systems with many parameters, it is more appropriate to work with a different figure of merit rather than with the covariance matrix. An important merit figure is the weighted mean squared error (WMSE). Consider that the estimator $\bar{t}$ is unbiased, that is, its expected value is in agreement with the parameter $\bar\theta=\bar\mu_{\bar\theta}$. In this case, the covariance matrix becomes the mean square error matrix (MSEM)
\begin{equation}
\Xi_{\bar\theta}\big(\boldsymbol{\Pi},\bar{t}\big)= \sum_{\omega\in\Omega} f(\omega|\bar\theta) \big(\bar{t}_{\bar\theta}(\omega)-\bar\theta\big)\big(\bar{t}_{\bar\theta}(\omega)-\bar{\theta}\big)^\top.
\end{equation}
The WMSE for the unbiased estimator $\bar{t}$ is defined by
\begin{equation}\label{RMSE}
w_{\bar{\theta}}\left(\boldsymbol{\Pi},\bar{t}\right) =\Tr\big[W_{\bar{\theta}}\Xi_{\bar{\theta}}\left(\boldsymbol{\Pi},\bar{t}\right)\big], 
\end{equation}
where $W_{\bar{\theta}}\in\R^{2k\times 2k}$ is a weighting matrix. It has been shown \cite{Suzuki2016,Fujiwara1999} that the following two quantities are both lower bounds of the WMSE,
\begin{align}
w^S_{\bar{\theta}}&=\Tr\left( W_{\bar{\theta}} ( J^S_{\bar\theta} )^{-1}   \right), \label{SBtoRMSE}\\
w^R_{\bar{\theta}}&= \Tr\left(W_{\bar{\theta}} \Real\left[(J_{\bar\theta}^R)^{-1}\right] \right) + \Tr\Abs\left( W_{\bar{\theta}}\Imag\left[\left(J_{\bar\theta}^R\right) ^{-1}\right]  \right), \label{RBtoRMSE}
\end{align}
where $\Abs(G)$ is defined  by
\begin{equation}
    \Abs(G)=O^{-1}\left[\begin{array}{ccc}|g_1|& & \\
    & \ddots &   \\
    & &|g_{n}|\end{array}\right] O,
\end{equation}
where $\{g_k\}_{k=1}^n$ are the eigenvalues of $G\in \mathbb{C}^{n\times n}$ and $O$ is the matrix that diagonalizes $G$. 

An important case is the estimation problem of parameters encoded in pure states. This problem is particularly relevant when the parameters to be estimated are associated with a family of unitary operations \cite{VanephQuantum2013}. It is well-known that the SLD-QFIM for a pure state model $\{\ket{\psi_{\bar{\theta}}} \}$ is given by
\begin{equation}
\big[J_{\bar{\theta}}^{S}\big]_{ij} = 4 \Real\bra{\partial_{\bar\theta_i}\psi_{\bar{\theta}}}\left(\mathbb{I}-\dyad{\psi_{\bar{\theta}}}{\psi_{\bar{\theta}}}\right)\ket{\partial_{\bar\theta_j}\psi_{\bar{\theta}}}\label{PureRealFisher},
\end{equation}
where $\bar{\theta}$ is the real representation of $\theta$ and $\ket{\partial_{\bar\theta_j}\psi_{\bar{\theta}}}:=\partial_{\bar\theta_j}\ket{\psi_{\bar{\theta}}}$. Notice that the SLD-QFIM is proportional to the Hessian of the Fubini-Study metric \cite{Stokes2020,Mari2021}
    \begin{equation}
        J^{S}_{\bar{\theta}} = \left.-2 \left[ \frac{\partial}{\partial \bar{\theta}} \left(\frac{\partial }{\partial\bar{\theta}} \vert \langle\psi_{\bar{\theta'}}\vert \psi_{\bar{\theta}} \rangle\vert^2 \right)^{\top} \right]\right|_{\bar{\theta'} = \bar{\theta}}.
    \end{equation}
Furthermore, a necessary and sufficient condition to saturate the SLD-QCRI of $\bar{t}$ for pure states is given by \cite{Matsumoto_2002,Ragy2016}
\begin{equation}\label{PureStateCondition}
\bra{\psi_{\bar{\theta}}}[ L^S_{\bar{\theta}_i}, L^S_{\bar{\theta}_j} ]\ket{\psi_{\bar{\theta}}} = 0,\qquad  i,j=1,\dots,k,
\end{equation}
with $[A,B]=AB-BA$ the commutator between $A$ and $B$. A particular case where this condition is satisfied is when all SLDs $\{L^S_{\bar{\theta}_i} \}$ commute. Thereby, all of them are diagonalizable on the same basis of eigenvectors, which correspond to the optimal measurement that saturates the QCRI.

It is not possible to define an RLD-QFIM for a low-rank state, such as a pure state, since \eqref{RQRLD} has no solution in this case \cite{Fujiwara1999}. However, for QCRB only the inverse of the QFIM is needed, which can be obtained approaching the pure state $\ket{\psi}$ by a family of full-rank mixed states $\rho(\epsilon)$ such that $\dyad{\psi}{\psi}=\lim_{\epsilon\rightarrow 0}\rho(\epsilon)$. To this end, in \cite{Holevo2011} (Chapter $6$) and \cite{Hayashi2010} (Chapter $18$) assumed that the commutation operator $\mathfrak{D}$ presented by Holevo in \cite{Holevo2011} is invariant on $\mathcal{T}_{\bar{\theta}}(\ket{\psi})$, the linear span (over $\mathbb{R}$) of the symmetric logarithmic derivatives \eqref{SLD}, and proved that 
\begin{equation} \label{RPureRealFisher}
\lim_{\epsilon \, \downarrow \, 0}\left(J_{\bar{\theta}}^R(\epsilon)\right)^{-1}=\left(J_{\bar{\theta}}^S\right)^{-1}+\rmi\left(J_{\bar{\theta}}^S\right)^{-1} K_{\bar{\theta}} \left(J_{\bar{\theta}}^S\right)^{-1}
\end{equation}
where 
\begin{equation}\label{KRRFIM}
\big[{K}_{\bar{\theta}}\big]_{ij} = 4\Imag \, \bra{\partial_{\bar{\theta}_i}\psi_{\bar\theta}}\left(\mathbb{I}-\dyad{\psi_{\bar{\theta}}}{\psi_{\bar{\theta}}}\right)\ket{{\partial_{\bar{\theta}_j}\psi_{\bar\theta}}},  
\end{equation}
$J_{\bar{\theta}}^R(\epsilon)$ is the RLD-QFIM for $\rho(\epsilon)$ and $J_{\bar{\theta}}^S$ is the SLD-QFIM for $\ket{\psi}$.


\section{Quantum estimation theory for complex parameters.}

The space state of quantum mechanics is typically formulated in the field of complex numbers. A quantum state provides all information required to predict the real-valued probabilities associated with the outcomes of any conceivable experiment, which is described through a generalized measurement. In general, any characteristics or information from a quantum state will be represented by a real number, which is a function of the complex coefficients that define a particular state. It seems thus convenient and consistent to formulate a quantum estimation theory that works directly with complex statistics depending on complex parameters. Here, we present an extension of the quantum estimation theory, briefly summarized in Section \ref{section_QET}, working on the field of complex numbers by means of the Wirtinger calculus.

We define the complex symmetric logarithm derivatives (CSLD) $\mathcal{L}^S_{\theta_i}$ and $\mathcal{L}^S_{\theta_i^*}$ of the state model $\rho_{\hat\theta}$ with respect to the complex parameters $\hat{\theta}$ and $\hat{\theta}^*$ as the solution of the following equations
\begin{equation}\label{CSLDeq}
\partial_{\theta_i^*}\rho =\frac{1}{2}\big(\rho_{\hat\theta} \mathcal{L}^S_{\theta_i^*}+\mathcal{L}^S_{\theta_i^*}\rho_{\hat\theta} \big),\qquad
 \partial_{\theta_i}\rho=\frac{1}{2}\big(\rho_{\hat\theta}\mathcal{L}^S_{\theta_i} + \mathcal{L}^S_{\theta_i}\rho_{\hat\theta} \big),
\end{equation}
respectively. Notice that, unlike the case of the real parameters, the CSLD are not hermitian. Instead, these satisfy the property $\mathcal{L}^S_{\theta_j} = {\mathcal{L}^S_{\theta_j^*}}^\dagger$. Using \eqref{Wirtinger derivatives} and \eqref{SLD}, we can show that there exists a relation between the CSLD and its real counterparts, that is,
\begin{equation}\label{CSLD}
\mathcal{L}^S_{\theta_i}=\frac{1}{2}\big(L^S_{\alpha_i}-\rmi L^S_{\beta_i}\big),\qquad\mathcal{L}^S_{\theta_i^*}=\frac{1}{2}\big(L^S_{\alpha_i}+\rmi L^S_{\beta_i}\big), \qquad i=1,\ldots, 2k.
\end{equation}
We define the CSLD-QFIM of state $\rho$ for the complex parameter $\hat \theta$ as
\begin{equation}\label{CSQFIM}
\mathcal{J}^S_{\hat\theta}=\left[\begin{array}{cc}\mathtt{J}^S_\theta & \mathtt{Q}^S_\theta\\
(\mathtt{Q}^S_\theta)^* & (\mathtt{J}_\theta^S)^*\end{array}\right],
\end{equation}
where the block matrices $\mathtt{J}_\theta^S$ and $\mathtt{Q}^S_\theta$ are the SLD-QFIM and the SLD pseudo-QFIM (pQFIM) of state $\rho_{\hat\theta}$ for the parameter $\theta$, given by 
\begin{equation}\label{csqfim}
\big[\mathtt{J}^S_{\theta}\big]_{ij}=\frac{1}{2}\Tr\Big(\rho_{\hat\theta} \big(\mathcal{L}^S_{\theta_i^*}\mathcal{L}^S_{\theta_j}+\mathcal{L}^S_{\theta_j}\mathcal{L}^S_{\theta_i^*}\big)\Big);\qquad
\big[\mathtt{Q}^S_{\theta}\big]_{ij}=\frac{1}{2}\Tr\Big(\rho_{\hat\theta} \big(\mathcal{L}^S_{\theta_i^*}\mathcal{L}^S_{\theta_j^*}+\mathcal{L}^S_{\theta_j^*}\mathcal{L}^S_{\theta_i^*}\big)\Big),
\end{equation}
for each $i,j=1,\ldots, k$. This result allows us to establish the following theorem:
\begin{teo}
Let $\bar{\theta}= [\alpha,\beta ]^{\top}$ be the real representation of $\theta$ and $\hat{\theta}= [\theta,\theta^* ]^{\top}$ its conjugate extension, $J^S_{\bar{\theta}}$ the SQFIM of state $\rho$ for the parameter $\bar \theta$, and $\mathcal{J}^S_{\hat\theta}$ the SQFIM of state $\rho$ for the parameter $\hat \theta$. Then
\begin{equation}\label{SQFIMequiv}
\mathcal{J}^S_{\hat\theta}=\frac{1}{4}\ev{J^S_{\bar{\theta}}}_\C. 
\end{equation}
\end{teo}

To prove this, we note that $J^S_{\bar\theta}$ can be expressed as a block matrix, that is,
\begin{equation*}
J^S_{\bar\theta}=\left[\begin{array}{cc}
J^S_{\alpha\alpha}& J^S_{\alpha\beta}\\ J^S_{\beta\alpha}& J^S_{\beta\beta}\end{array}\right],
\end{equation*}
where $[J^S_{ab}]_{ij}=\Tr\big(\rho_{\hat\theta}[L^S_{a_i}L^S_{b_j}+L^S_{b_j}L^S_{a_i}]\big)/2$ and $a,b\in\{\alpha,\beta\}$. Then, using the map $\ev{\cdot}_\C$ we obtain
\begin{equation}\label{QFIMBlock2}
\ev{J_{\bar\theta}}_\C=\left[\begin{array}{cc}J^S_{\alpha\alpha}+\rmi J^S_{\beta\alpha}-\rmi(J^S_{\alpha\beta}+\rmi J_{\beta\beta}) & J^S_{\alpha\alpha}+\rmi J^S_{\beta\alpha}+\rmi(J^S_{\alpha\beta}+\rmi J^S_{\beta\beta})\\
J^S_{\alpha\alpha}-\rmi J^S_{\beta\alpha}-\rmi(J^S_{\alpha\beta}-\rmi J^S_{\beta\beta})&
J^S_{\alpha\alpha}-\rmi J^S_{\beta\alpha}+\rmi(J^S_{\alpha\beta}-\rmi J^S_{\beta\beta})\end{array}\right],
\end{equation}
where, from \eqref{CSLD}, the first block is given by
\begin{align}
 \big[J^S_{\alpha\alpha}+\rmi J^S_{\beta\alpha}-\rmi(J^S_{\alpha\beta}+\rmi J^S_{\beta\beta})\big]_{ij}&=\frac{1}{2}\Tr\Big(\rho_{\hat\theta} \big[(L^S_{\alpha_i}+\rmi L^S_{\beta_i})(L^S_{\alpha_j}-\rmi L^S_{\beta_j})\\
&\qquad+(L^S_{\alpha_j}-\rmi L^S_{\beta_j})(L^S_{\alpha_i}+\rmi L^S_{\beta_i})\big]\Big),\\
&=\frac{1}{2}\Tr\Big(\rho_{\hat\theta} \big(\mathcal{L}^S_{\theta_i^*}\mathcal{L}^S_{\theta_j}+\mathcal{L}^S_{\theta_j}\mathcal{L}^S_{\theta_i^*}\big)\Big).
\end{align}
Similarly, with the remaining blocks, we obtain
\begin{equation*}
 \ev{J^S_{\bar\theta}}_\C=4\left[\begin{array}{cc}\frac{1}{2}\Tr\Big(\rho_{\hat\theta} \big(\mathcal{L}^S_{\theta_i^*}\mathcal{L}^S_{\theta_j}+\mathcal{L}^S_{\theta_j}\mathcal{L}^S_{\theta_i^*}\big)\Big) & 
\frac{1}{2}\Tr\Big(\rho_{\hat\theta} \big(\mathcal{L}^S_{\theta_i^*}\mathcal{L}^S_{\theta_j^*}+\mathcal{L}^S_{\theta_j^*}\mathcal{L}^S_{\theta_i^*}\big)\Big)\\
\frac{1}{2}\Tr\Big(\rho_{\hat\theta} \big(\mathcal{L}^S_{\theta_i^*}\mathcal{L}^S_{\theta_j^*}+\mathcal{L}^S_{\theta_j^*}\mathcal{L}^S_{\theta_i^*}\big)\Big)^*&
\frac{1}{2}\Tr\Big(\rho_{\hat\theta} \big(\mathcal{L}^S_{\theta_i^*}\mathcal{L}^S_{\theta_j}+\mathcal{L}^S_{\theta_j}\mathcal{L}^S_{\theta_i^*}\big)\Big)^*\end{array}\right],
\end{equation*}
and conclude the proof of \eqref{CSQFIM} and  \eqref{csqfim}.


The above analysis can also be applied to obtain a complex version of the RLD-QFIM. We define the complex right logarithm derivatives (CRLD) $\mathcal{L}_{\theta_i}^R$ and $\mathcal{L}_{\theta_i^*}^R$ as the solutions of the following equations
\begin{equation}
\partial_{\theta_i^*}\rho=\rho_{\hat\theta}\mathcal{L}^R_{\theta_i^*};\qquad
\partial_{\theta_i}\rho=\rho_{\hat\theta}\mathcal{L}^R_{\theta_i},\qquad i=1,\ldots, k,
\end{equation}
respectively. Notice that CRLD $\mathcal{L}^R_{\theta_i}$ are not hermitian. In addition, using \eqref{Wirtinger derivatives} and \eqref{RQRLD}, we can obtain a relation between the CRLD and its real counterparts 
\begin{equation}\label{CRLD}
\mathcal{L}^R_{\theta_i}=\frac{1}{2}\big(L^R_{\alpha_i}-\rmi L^R_{\beta_i}\big),\qquad\mathcal{L}^R_{\theta_i^*}=\frac{1}{2}\big(L^R_{\alpha_i}+\rmi L^R_{\beta_i}\big),\qquad i=1,\ldots, k.
\end{equation}
Motivated by \eqref{CCFIM} and \eqref{RRQFIM}, we define the CRLD-QFIM of state $\rho$ for the complex parameters $\hat\theta$ as
\begin{equation}\label{CRQFIM}
\mathcal{J}^R_{\hat\theta}=\left[\begin{array}{cc}\mathtt{J}^R_\theta&\mathtt{Q}^R_\theta\\
\mathtt{Q}^R_{\theta^*}&\mathtt{J}_{\theta^*}^R\end{array}\right],
\end{equation}
where the RLD-QFIM and the RLD-pQFIM of state $\rho$ for the parameter $\theta$ are given by
\begin{equation}\label{crqfim}
\big[\mathtt{J}^R_{\theta}\big]_{ij}=\Tr\Big(\rho_{\hat\theta} \Big[\mathcal{L}^R_{\theta_j}{\mathcal{L}^R_{\theta_i}}^\dagger\Big]\Big),\qquad
\big[\mathtt{Q}^R_{\theta}\big]_{ij}=\Tr\Big(\rho_{\hat\theta} \Big[\mathcal{L}^R_{\theta_j^*}{\mathcal{L}^R_{\theta_i}}^\dagger\Big]\Big),
\end{equation}
for each $i,j=1,\ldots, k$. Now, we can establish the following theorem.
\begin{teo}
Let $\bar{\theta}$ be the real representation of $\theta$ and $\hat{\theta}$ its conjugate extension, $J^R_{\bar{\theta}}$  the RQFIM of $\rho_{\hat{\theta}}$ for the parameter $\bar \theta$ and $\mathcal{J}^R_{\hat\theta}$ the RQFIM of $\rho_{\hat{\theta}}$ for the parameter $\hat \theta$. Then
\begin{equation}\label{RQFIMequiv}
\mathcal{J}^R_{\hat\theta}=\frac{1}{4}\ev{J^R_{\bar{\theta}}}_\C.
\end{equation}
\end{teo}

To prove this, we consider the block matrix
\begin{equation*}
J^R_{\bar\theta}=\left[\begin{array}{cc}J^R_{\alpha\alpha} & J^R_{\alpha\beta}\\ J^R_{\beta\alpha} & J^R_{\beta\beta}\end{array}\right],
\end{equation*}
where $[J^R_{ab}]_{ij}=\Tr\Big(\rho_{\hat\theta} L^R_{b_j}{L^R_{a_i}}^\dagger\Big)$ and $a,b\in \{\alpha,\beta\}$. Then applying the map $\ev{\cdot}_\C$, we have that
\begin{equation}\label{mapRRQFIM}
 \ev{J^R_{\bar\theta}}_\C=\left[\begin{array}{cc}J^R_{\alpha\alpha}+\rmi J^R_{\beta\alpha}-\rmi(J^R_{\alpha\beta}+\rmi J^R_{\beta\beta}) & J^R_{\alpha\alpha}+\rmi J^R_{\beta\alpha}+\rmi(J^R_{\alpha\beta}+\rmi J^R_{\beta\beta})\\
J^R_{\alpha\alpha}-\rmi J^R_{\beta\alpha}-\rmi(J^R_{\alpha\beta}-\rmi J^R_{\beta\beta})&
J^R_{\alpha\alpha}-\rmi J^R_{\beta\alpha}+\rmi(J^R_{\alpha\beta}-\rmi J^R_{\beta\beta})\end{array}\right],
\end{equation}
working on the first block of \eqref{mapRRQFIM} and recalling \eqref{CRLD}, we see that its components are given by
\begin{align}
\big[J^R_{\alpha\alpha}+\rmi J^R_{\beta\alpha}-\rmi(J^R_{\alpha\beta}+\rmi J^R_{\beta\beta})\big]_{ij}&=\Tr\Big(\rho_{\hat\theta} \big[(L^R_{\alpha_j}-\rmi L^R_{\beta_j})(L^R_{\alpha_i}-\rmi L^R_{\beta_i})^\dagger\big]\Big),\\
&=4\Tr\Big(\rho_{\hat\theta}\Big(\mathcal{L}^R_{\theta_j}{\mathcal{L}^R_{\theta_i}}^\dagger\Big)\Big).
\end{align}
Analogously to the other blocks, we conclude that
\begin{equation*}
\ev{J^R_{\bar\theta}}_\C=4\left[\begin{array}{cc}\Tr\Big(\rho_{\hat\theta} \Big(\mathcal{L}^R_{\theta_j}{\mathcal{L}^R_{\theta_i}}^\dagger\Big)\Big) &
\Tr\Big(\rho_{\hat\theta} \Big(\mathcal{L}^R_{\theta_j^*}{\mathcal{L}^R_{\theta_i}}^\dagger\Big)\Big)\\
\Tr\Big(\rho_{\hat\theta} \Big(\mathcal{L}^R_{\theta_j}{\mathcal{L}^R_{\theta_i^*}}^\dagger\Big)\Big) &
\Tr\Big(\rho_{\hat\theta} \Big(\mathcal{L}^R_{\theta_j^*}{\mathcal{L}^R_{\theta_i^*}}^\dagger\Big)\Big)\end{array}\right],
\end{equation*}
which finishes the proof of \eqref{CRQFIM} and \eqref{crqfim}.


Having defined CSLD-QFIM and CRLD-QFIM, we can state the quantum Cramér-Rao inequality (QCRI) for complex parameters as follows:
\begin{teo}
Let $\hat \theta$ be the conjugate extension of state $\theta$, $\mathcal{I}_{\hat\theta}$ the classical FIM of $\rho$ for the parameter $\hat{\theta}$, and $\mathcal{J}^X_{\hat\theta}$ the CSLD-QFIM or CRLD-QFIM of state $\rho_{\hat\theta}$ for the parameter $\hat{\theta}$. Then
\begin{equation}\label{FisherMatricesInequality}
\mathcal{I}_{\hat\theta}\leq \mathcal{J}^X_{\hat\theta},\qquad X \in \{S,R\}.
\end{equation}
Morever, let $\Cov_{\hat{\theta}}(\boldsymbol{\Pi},\hat{t})$ be the covariance matrix and $\mathcal{D}_{\hat\mu}$ the Wirtinger Jacobian matrix, then 
\begin{equation}\label{CQCRB}
\Cov_{\hat{\theta}}(\boldsymbol{\Pi},\hat{t}) \geq\mathcal{D}_{\hat\mu} \mathcal{I}_{\hat\theta}^{-1}\mathcal{D}_{\hat\mu}^\dagger\geq\mathcal{D}_{\hat\mu} \left(\mathcal{J}^X_{\hat\theta}\right)^{-1}\mathcal{D}_{\hat\mu}^\dagger.
\end{equation}
\end{teo}

To obtain \eqref{FisherMatricesInequality}, we apply the map $\ev{\cdot}_\C$ to \eqref{QCRB} and use the definitions \eqref{CCFIM}, \eqref{RSQFIM} and \eqref{RRQFIM}. 

The real and complex QCRI are equivalent, that is, 
\begin{equation}\label{CQCRBequiv} 
\Cov_{\bar{\theta}}\left(\bar{t}\right)\geq D_{\bar{\mu}}\left(J^X_{\bar\theta}\right)^{-1}D_{\bar{\mu}}^\top \Longleftrightarrow \Cov_{\hat{\theta}}\left(\hat{t}\right)\geq\mathcal{D}_{\hat\mu}\left(\mathcal{J}^X_{\hat\theta}\right)^{-1}\mathcal{D}_{\hat\mu}^\dagger,
\end{equation}
and both are simultaneously attained, that is, if $\bar{t}$ reaches the real QCRB, then $\hat{t}$ reaches the complex QCRB and vise versa. 

The relation \eqref{CQCRB} provides a compact bound for the covariance of $\hat{t}$, which includes a bound for $t$. However, to obtain an explicit bound for the covariance of $t$, that is,
\begin{equation} \label{covariance_theta}
\Cov_{\theta}(\boldsymbol{\Pi},t) = \sum_{\omega\in\Omega} f(\omega|\theta) \big(t_{\theta}(\omega)-\mu_{\theta}\big)\big(t_{\theta}(\omega)-\mu_{\theta}\big)^\dagger, 
\end{equation}
with $\mu_{\theta}= \sum_{\omega\in\Omega} f(\omega|\theta) t_{\theta}(\omega)$ the expected value of $t$, we have to work with the block form of the QFIM $\mathcal{J}^X_{\hat\theta}$, with $X$ in $\{S,R\}$.
\begin{cor}
Let $\hat{\theta}$ be the conjugate extension of $\theta$, and $\mathcal{J}^S_{\hat\theta}$ the CSLD-QFIM of state $\rho_{\hat{\theta}}$ for the parameter $\hat{\theta}$ given by \eqref{CSQFIM} and \eqref{csqfim}, and considering
\begin{equation}
 \left(\mathcal{J}^S_{\hat\theta}\right)^{-1}=\left[\begin{array}{cc}\mathtt{J}^S_\theta &\mathtt{Q}^S_\theta \\ \left(\mathtt{Q}^S_{\theta}\right)^* & \left(\mathtt{J}^S_{\theta}\right)^*\end{array}\right]^{-1}
=\left[\begin{array}{cc}\left(\mathtt{E}_\theta^S\right)^{-1} & -\mathtt{F}_\theta^S\left(\mathtt{E}_\theta^S\right)^{-1}\\
-\left(\mathtt{F}_\theta^S\right)^*\left(\mathtt{E}_\theta^S\right)^{-*} & \left(\mathtt{E}_\theta^S\right)^{-*}\end{array}\right], \label{BLockCQFIM}
\end{equation}
where $\mathtt{E}_\theta^S=\mathtt{J}^S_\theta-\mathtt{Q}^S_\theta\big(\mathtt{J}_{\theta}^{S}\big)^{-*}\left(\mathtt{Q}^S_{\theta}\right)^*$ and $\mathtt{F}^S_\theta=\big(\mathtt{J}_{\theta}^{S}\big)^{-1}\mathtt{Q}_\theta^S$, and $(\cdot)^{-*}$ denotes the conjugate inverse. Then
\begin{equation}\label{ICSQIMBF}
\begin{aligned}
\Cov_\theta \left(\boldsymbol{\Pi}, t\right)\geq & \; \mathcal{D}_\theta \mu_\theta \left(\mathtt{E}_\theta^S\right)^{-1} \left(\mathcal{D}_{\theta} \mu_\theta\right)^\dagger - \mathcal{D}_{\theta^*} \mu_\theta \left(\mathtt{F}_\theta^S\right)^*\left(\mathtt{E}_\theta^S\right)^{-*} \left(\mathcal{D}_{\theta} \mu_\theta \right)^\dagger\\
&-\mathcal{D}_\theta \mu_\theta \mathtt{F}_\theta^S\left(\mathtt{E}_\theta^S\right)^{-1} \left(\mathcal{D}_{\theta^*} \mu_\theta \right)^\dagger + \mathcal{D}_{\theta^*}\mu_\theta \left(\mathtt{E}_\theta^S\right)^{-1} \left(\mathcal{D}_{\theta^*} \mu_\theta\right)^\dagger.
\end{aligned}
\end{equation}
\end{cor}
An analog result can be obtained for the right logarithmic derivative as follows:
\begin{cor}
Let $\hat{\theta}$ be the conjugate extension of $\theta$, and $\mathcal{J}^R_{\hat\theta}$ the CRLD-QFIM of state $\rho_{\hat\theta}$ for the parameter $\hat{\theta}$ given by \eqref{CRQFIM} and \eqref{crqfim}, and considering
\begin{equation}
\left(\mathcal{J}^R_{\hat\theta}\right)^{-1}=\left[\begin{array}{cc}\mathtt{J}^R_\theta&\mathtt{Q}^R_\theta \\
\mathtt{Q}^R_{\theta^*} & \mathtt{J}^R_{\theta^*}\end{array}\right]^{-1}=\left[\begin{array}{cc}\left(\mathtt{E}_\theta^R\right)^{-1}&-\mathtt{F}_\theta^R\left(\mathtt{E}_\theta^R\right)^{-1}\\[0.1 cm] -\mathtt{F}_{\theta^*}^R\left(\mathtt{E}_{\theta^*}^R\right)^{-1}&\left(\mathtt{E}_{\theta^*}^R\right)^{-1}\end{array}\right], \label{BLockCQFIM2}
\end{equation}
where $\mathtt{E}_{\theta}^R=\mathtt{J}^R_\theta-\mathtt{Q}^R_\theta\big(\mathtt{J}_{\theta^*}^{R}\big)^{-1}\mathtt{Q}^R_{\theta^*}$ and $\mathtt{F}^R_\theta=\mathtt{Q}_\theta^R\big(\mathtt{J}_{\theta}^{R}\big)^{-1}$. Then
\begin{equation}\label{ICRQIMBF}
\begin{aligned}
\Cov_\theta\left(\boldsymbol{\Pi}, t\right)\geq & \;\mathcal{D}_\theta \mu_\theta \left(\mathtt{E}_\theta^R\right)^{-1} \left(\mathcal{D}_{\theta} \mu_\theta \right)^\dagger - \mathcal{D}_{\theta^*} \mu_\theta \mathtt{F}_{\theta^*}^R\left(\mathtt{E}_{\theta^*}^R\right)^{-1} \left(\mathcal{D}_{\theta} \mu_\theta \right)^\dagger\\
&-\mathcal{D}_\theta \mu_\theta \mathtt{F}_\theta^R\left(\mathtt{E}_\theta^R\right)^{-1} \left(\mathcal{D}_{\theta^*} \mu_\theta \right)^\dagger + \mathcal{D}_{\theta^*} \mu_\theta \left(\mathtt{E}_{\theta^*}^R\right)^{-1} \left(\mathcal{D}_{\theta^*} \mu_\theta \right)^\dagger.
\end{aligned}
\end{equation}
\end{cor}
Unlike the symmetric case, the matrices $\mathtt{E}_\theta^R $ and $\mathtt{E}_{\theta^*}^R $ are independent. In addition, in both cases, when the pQFIM $\mathtt{Q}^X_{\theta^*}$ vanishes, the matrices $(\mathcal{J}^X_{\hat\theta})^{-1}$ become block diagonal, and in addition, if $t$ is an unbiased estimator, then the inequality is simplified to
\begin{equation}
\Xi_\theta\left(\boldsymbol{\Pi},t\right)\geq\left(\mathtt{J}_\theta^X\right)^{-1}, \qquad X\in \{S,R\},
\end{equation}
with $\Xi_\theta\left(\boldsymbol{\Pi},t\right)$ the MSEM of the estimator $t$ for the parameter $\theta$, obtained replacing $\mu_\theta = \theta$ is the equation \eqref{covariance_theta}.


To obtain a lower bound for the weighted mean square error $w_{\hat\theta}$ in the complex parameter case, we define \begin{equation}\label{CMSE}
w_{\hat\theta}\big(\hat{t}\big) = \Tr\left(\mathcal{W}_{\hat{\theta}}\Cov_{\hat\theta}\left(\hat{t}\right) \right),
\end{equation}
where $\mathcal{W}_{\hat{\theta}}$ is the complex weighting matrix for the complex parameter $\hat{\theta}$ defined as
\begin{equation}\label{CWMBF}
\mathcal{W}_{\hat{\theta}}=\left[\begin{array}{cc}\mathtt{W}_\theta& \mathtt{X}_\theta\\ (\mathtt{X}_\theta)^*&(\mathtt{W}_\theta)^*\end{array}\right],
\end{equation}
with complex weighting matrices $\mathtt{W}_\theta$ and $\mathtt{X}_\theta$ for the complex parameter $\theta$. In many cases, it is convenient to consider $\mathcal{W}_{\hat{\theta}}$ as a real matrix, but it may be useful to consider it complex to study the RLD limit \eqref{RBtoRMSE}.

\begin{teo}
Let $\bar{t}$ be the real transformation of $t$, and $\hat{t}$ its conjugate extension. If the real and complex weighting matrices fulfill  
\begin{equation}
\mathcal{W}_{\hat{\theta}}=\frac{1}{4}\ev{W_{\bar{\theta}}}_{\mathbb{C}},
\label{CWM}
\end{equation}
then
\begin{equation}
w_{\hat\theta}(\hat{t})=w_{\bar\theta}(\bar{t}).
\end{equation}
\end{teo}

To prove this, we use \eqref{COV equiv} and the property $\Tr(AB)=\Tr(BA)$ for complex matrices $A$ and $B$. From this result, we find that \eqref{SBtoRMSE} and \eqref{RBtoRMSE} are also lower bounds for $w_{\hat\theta}(\hat{t})$.


\begin{teo}
Let $\hat \theta$ be the conjugate extension of $\theta$, $w^S_{\hat\theta}$ and $w^R_{\hat\theta}$ be the bounds to $w_{\hat\theta}$ defined in 
\eqref{SBtoRMSE} and \eqref{RBtoRMSE}, $\mathcal{W}_{\hat{\theta}}$ the complex weighting matrix satisfying \eqref{CWM}, $\mathcal{J}^S_{\hat{\theta}}$ and $\mathcal{J}^R_{\hat{\theta}}$ the CSLD- and CRLD-QFIM of state $\rho_{\hat\theta}$ for the parameter $\hat \theta$. Then
\begin{equation}
w_{\hat\theta}^S=\Tr\left(\mathcal{W}_{\hat{\theta}}\left(\mathcal{J}^S_{\hat{\theta}}\right)^{-1}\right),\label{SBtoCMSE}
\end{equation}\vspace{-0.25cm}
\begin{equation}
\begin{aligned}
w_{\hat\theta}^R=& \frac{1}{2} \Tr\left(\mathcal{W}_{\hat{\theta}}\left[(\mathcal{J}_{\hat{\theta}}^R)^{-1}+\sigma_{2k}(\mathcal{J}_{\hat{\theta}}^R)^{-*}\sigma_{2k}\right] \right)\\
&+ \frac{1}{2} \Tr\Abs\left(\mathcal{W}_{\hat{\theta}}\left[(\mathcal{J}_{\hat\theta}^R)^{-1} -\sigma_{2k}(\mathcal{J}_{\hat\theta}^R)^{-*}\sigma_{2k}\right] \right).\label{RBtoCMSE}
\end{aligned}
\end{equation}
\end{teo}

To prove \eqref{SBtoCMSE}, we use \eqref{CWM} and apply the trace property \eqref{property7} of the map $\ev{\cdot}_\C$. On the other hand, to prove \eqref{RBtoCMSE}, we note that using properties \eqref{property4}, \eqref{property7} and \eqref{property9}, and \eqref{RQFIMequiv} and \eqref{CWM} in the first term of \eqref{RBtoRMSE}, we obtain
\begin{equation}\label{first term}
\begin{aligned}
\Tr\left(W_{\bar{\theta}}\left( (J_{\bar\theta}^R)^{-1}+(J_{\bar\theta}^R)^{-*}\right) \right)
&=\Tr\left(\ev{W_{\bar{\theta}}}_\C\left( \ev{J_{\bar\theta}^R}_\C^{-1} + \ev{(J_{\bar\theta}^R)^{*}}_\C^{-1}\right) \right),\\
&=\Tr\left(\mathcal{W}_{\hat{\theta}}\left[(\mathcal{J}_{\hat{\theta}}^R)^{-1} +\sigma_{2k}(\mathcal{J}_{\hat{\theta}}^R)^{-*}\sigma_{2k}\right] \right).
\end{aligned}
\end{equation}
In addition, using properties \eqref{property9},\eqref{property7}, \eqref{property2}, \eqref{property4}, and \eqref{property5}, and \eqref{RQFIMequiv} and \eqref{CWM} in the second term of \eqref{RBtoRMSE}, we have the that
\begin{equation}\label{second term}
\begin{aligned}
\Tr\Abs\left(W_{\bar{\theta}}\left[ (J_{\bar\theta}^R)^{-1}-(J_{\bar\theta}^R)^{-*}\right] \right)&=\frac{1}{2}\Tr\Abs\left(\ev{W_{\bar{\theta}}\left[ (J_{\bar\theta}^R)^{-1}-(J_{\bar\theta}^R)^{-*}\right]}_\C \right),\\
&=\Tr\Abs\left(  \ev{W_{\bar{\theta}}}_\C\left[ \ev{J_{\bar\theta}^R}_\C^{-1}-\ev{(J_{\bar\theta}^R)^{*}}_\C^{-1}\right] \right),\\
&=\Tr\Abs\left(\mathcal{W}_{\hat{\theta}}\left[(\mathcal{J}_{\hat\theta}^R)^{-1} -\sigma_{2k}(\mathcal{J}_{\hat\theta}^R)^{-*}\sigma_{2k}\right] \right),
\end{aligned}
\end{equation}
which allows us to complete the proof by replacing \eqref{first term} and \eqref{second term} in \eqref{RBtoRMSE}.

Bounds \eqref{SBtoCMSE} and \eqref{RBtoCMSE} can be explicitly written considering the block forms of $\mathcal{W}_{\hat{\theta}}$ and $\mathcal{J}_{\hat\theta}^X$. So,
\begin{equation}
 w^S_\theta = 2\Tr\left(\Real\left(\mathtt{W}_\theta\left(\mathtt{E}_\theta^S\right)^{-1} +\mathtt{X}_\theta\left(\mathtt{F}_\theta^S\right)^*\left(\mathtt{E}_\theta^S\right)^{-*}  \right)\right),
\end{equation}
\begin{equation}
\begin{aligned}
w^R_\theta & =   \Tr\left(\Real\left( \mathtt{W}_\theta\left(\mathtt{E}_\theta^R\right)^{-1}  +\mathtt{W}_\theta^*\left(\mathtt{E}_{\theta^*}^R\right)^{-1} -\mathtt{X}_\theta\mathtt{F}_{\theta^*}^R\left(\mathtt{E}_{\theta^*}^R\right)^{-1} -\mathtt{X}^*_\theta\mathtt{F}_{\theta}^R\left(\mathtt{E}_{\theta}^R\right)^{-1}    \right)\right)\\
 & + \Tr \Abs \thinspace \left(\Imag\left( \mathtt{W}_\theta\left(\mathtt{E}_\theta^R\right)^{-1}+\mathtt{W}_\theta^*\left(\mathtt{E}_{\theta^*}^R\right)^{-1} -  \mathtt{X}_\theta \mathtt{F}_{\theta^*}^S\left(\mathtt{E}_{\theta^*}^R\right)^{-1}  -\mathtt{X}^*_\theta\mathtt{F}_{\theta}^R\left(\mathtt{E}_{\theta}^R\right)^{-1}\right)\right).
 \end{aligned}
\end{equation}
Furthermore, when pQFIM $\mathtt{Q}_\theta^X$ is null, for $X\in\{S,R\}$, the WMSE bounds become
\begin{equation}
w^S_{\theta}=2\Tr\left(\Real\left(\mathtt{W}_\theta (\mathtt{J}_\theta^S)^{-1}\right)\right),
\end{equation}
\begin{equation}
\begin{aligned}
w^R_{\theta}=&\Tr\left(\Real\left(\mathtt{W}_\theta(\mathtt{J}^R_\theta)^{-1}+\mathtt{W}^*_\theta(\mathtt{J}^R_{\theta^*})^{-1}\right)\right)+\Tr \Abs\thinspace \left( \Imag \left(\mathtt{W}_\theta(\mathtt{J}^R_\theta)^{-1}+ \mathtt{W}^*_\theta(\mathtt{J}^R_{\theta^*})^{-1}\right)\right).
\end{aligned}
\end{equation}


Finally, we specialize our previous results to the case of pure states. We denote the Wirtinger derivative of the pure states as $\ket{\partial_{z}\psi}=\partial_z\ket{\psi}$ for $z\in\mathbb{C}$. Let us note that $\ket{\partial_{z}\psi}^\dagger=\bra{\partial_{z^*}\psi}$.
\begin{teo}
	Let $\hat{\theta}$ be the conjugate extension of $\theta$, and $\mathcal{J}_{\hat \theta}^S$ the CSLD-QFIM of state $\ket{\psi_{\hat\theta}}$ for the parameter $\hat{\theta}$. If we consider $\mathcal{J}_{\hat \theta}^S$ in its block form \eqref{CSQFIM}, then its components are given by
	\begin{equation}\label{CFM1}
	\left[\mathtt{J}_{\theta}^S\right]_{jk}
    = 4\, \Real\left[\bra{\partial_{\theta^*_j}\psi_{\hat\theta}}\left(\mathbb{I}-\dyad{\psi_{\hat\theta}}{\psi_{\hat\theta}}\right)\ket{\partial_{\theta_k}\psi_{\hat\theta}} \right],
	\end{equation}
	\begin{equation}\label{CFM2}
    [\mathtt{Q}^S_{\theta}]_{jk} = 4 \, \Real\left[\bra{\partial_{\theta^*_j}\psi_{\hat\theta}}\left(\mathbb{I}-\dyad{\psi_{\hat\theta}}{\psi_{\hat\theta}}\right)\ket{\partial_{\theta^*_k}\psi_{\hat\theta}} \right].
	\end{equation}
\end{teo}

To obtain \eqref{CFM1}, we employ \eqref{PureRealFisher} in \eqref{QFIMBlock2} and use \eqref{CSQFIM} and \eqref{SQFIMequiv}. Similarly, we obtain \eqref{CFM2}.

For the above result, we note that
\begin{equation}
\mathcal{J}^{S}_{\hat{\theta}} = \left.-2 \left[ \frac{\partial}{\partial \hat{\theta}} \left(\frac{\partial \vert \langle\psi_{\hat{\theta'}} \vert \psi_{\hat{\theta}} \rangle\vert^2}{\partial\hat{\theta}}\right)^{\dagger} \right]\right|_{\hat{\theta'} = \hat{\theta}}.
\end{equation}
Besides, when the state $\ket{\psi_{\hat\theta}}$ is only a function of $\theta$, and not of $\theta^*$, the pQFIM $\mathtt{Q}^S_{\theta}$ is identically zero.

\begin{teo}
	Let $\ket{\psi_{\hat\theta}}$ be a pure state model depending on complex parameters $\theta$. Let $\hat{\theta}$ be the conjugate extensions of $\theta$, and $\{L^S_{\hat{\theta}_j} \}$ the CSLDs of $\ket{\psi_{\hat\theta}}$. A necessary and sufficient condition to attain the SLD quantum Cram\'er-Rao bound, that is, $\Cov(\hat{t}) = (\mathcal{J}_{\hat\theta}^S)^{-1}$, is
	\begin{equation}
	\bra{\psi_{\hat\theta}}[ \mathcal{L}_{\hat{\theta}_j}^S, \mathcal{L}_{\hat{\theta}_k}^S ]\ket{\psi_{\hat\theta}} = 0. \label{QCRB_opt_cond}
	\end{equation}
\end{teo}

To prove this, we replace \eqref{CSLD} in \eqref{PureStateCondition}.

To compute $(\mathcal{J}_{\hat \theta}^R)^{-1}$ for a pure state $\ket{\psi}$, this must be set as a limit value of a sequence of full-rank mixing states as in \eqref{RPureRealFisher}. So, we define by $\mathcal{T}_{\hat{\theta}}(\ket{\psi})$ the linear span (over $\mathbb{R}$) of the complex symmetric logarithmic derivatives \eqref{CSLD}. Besides, we define
\begin{equation}\label{KRCFIM}
\mathcal{K}_{\hat\theta} = \left[ \begin{array}{cc} \mathtt{K}_\theta & \mathtt{R}_\theta \\ \mathtt{R}_\theta^*& \mathtt{K}_\theta^* \end{array} \right],
\end{equation}
where
\begin{equation}\label{RFIM1_pure}
\mathtt{K}_\theta = 4 \, \Imag \bra{\partial_{\theta^*_j}\psi_{\hat\theta}}\left(\mathbb{I}-\dyad{\psi_{\hat\theta}}{\psi_{\hat\theta}}\right)\ket{\partial_{\theta_k}\psi_{\hat\theta}},
\end{equation}
\begin{equation}\label{RFIM2_pure}
\mathtt{R}_{\theta} =  4 \, \Imag \bra{\partial_{\theta^*_j}\psi_{\hat\theta}}\left(\mathbb{I}-\dyad{\psi_{\hat\theta}}{\psi_{\hat\theta}}\right)\ket{\partial_{\theta^*_k}\psi_{\hat\theta}}.
\end{equation}
This matrix is related to its real counterpart in \eqref{RPureRealFisher} by
\begin{equation}\label{CK}
\mathcal{K}_{\hat\theta}=\frac{1}{4}\ev{K_{\bar{\theta}}}_\C, 
\end{equation}
and allow us to enunciate the following theorem:

\begin{teo}
	Let $\hat{\theta}$ the conjugate extension of $\theta$, $\mathcal{J}_{\hat\theta}^R$ the CRLD-QFIM of state $\ket{\psi_{\hat\theta}}$ for the complex parameter $\hat{\theta}$, and $\left\{\rho_{\hat\theta}(\epsilon): \epsilon > 0\right\}$ a family of strictly positive density operators which satisfy $\lim_{\epsilon\downarrow 0}= \ket{\psi}\bra{\psi}$. If $\mathcal{T}_{\hat\theta}(\ket{\psi})$ is $\mathfrak{D}$-invariant, then
	\begin{equation}\label{ComplexPureQFIM}
	\lim_{\epsilon\, \downarrow \, 0}\left(\mathcal{J}_{\hat\theta}^R\right)^{-1}(\epsilon)=\left(\mathcal{J}_{\hat\theta}^S\right)^{-1}+\rmi\left(\mathcal{J}_{\hat\theta}^S\right)^{-1} \mathcal{K}_{\hat\theta}\left(\mathcal{J}_{\hat\theta}^S\right)^{-1}.
	\end{equation}
\end{teo}

This result is proved by a similar procedure illustrated in \cite{Holevo2011} (Chapter $6$) and \cite{Hayashi2005} (Chapter $18$).


\section{Estimation of complex parameter encoded on coherent states}

Complex parameter estimation plays a critical role in many quantum protocols. Consequently, our proposal could have a wide range of applications. Next, we present an example of the application of the complex-field formulation of quantum estimation theory. Let us consider the problem of estimating a complex number $z$ encoded in a coherent state of the electromagnetic field, that is,
\begin{equation}\label{coherentstate}
|\alpha\rangle = D(\alpha)|0\rangle,
\end{equation}
where $D(\alpha)= \exp(\alpha a^\dagger - \alpha^* a)$ is the displacement operator, $a$ and $a^{\dagger}$ are the anihilation and creation operators. The complex number $z$ is encoded in the parameter $\alpha$ of the coherent state as
\begin{equation}
\alpha_{\hat z} = \epsilon z + \eta z^*,
\end{equation}
where $(\epsilon,\eta)$ are known complex parameters that play the role of a key. We are interested in the measurement set that yields $z$ with the smallest possible uncertainty. This situation arises in quantum communication, where a sender encodes a message $z$ in a coherent state $\ket{\alpha_{\hat z}}$ and later sends this state through a physical channel to a receiver, who must decode the message. The optimal coding/decoding scheme can be obtained using the real-field formulation of estimation theory \cite{FujiwaraCoherent1999, Arnhem2019}, but here we solve it using our proposal.

From \eqref{QCRB_opt_cond}, we have that CSLD-QCRB is attained if
\begin{equation}\label{QCRB_opt_cond_example}
\langle\alpha_{\hat z}|[ \mathcal{L}_{z}^S, \mathcal{L}_{z^*}^S ]|\alpha_{\hat z}\rangle = 0, 
\end{equation}
where $\mathcal{L}_{z}^S$ are the CSLD of $|\alpha\rangle$, which are implicitly given by \eqref{CSLDeq}, that is,
\begin{equation}
\partial_z(|\alpha_{\hat z}\rangle\langle\alpha_{\hat z}| ) = \frac{1}{2}(|\alpha_{\hat z}\rangle\langle\alpha_{\hat z}|\mathcal{L}_{z}^S + \mathcal{L}_{z}^S|\alpha_{\hat z}\rangle\langle\alpha_{\hat z}|  ).
\end{equation}
To compute $\mathcal{L}_z^S$ we must calculate the Wirtinger derivatives of $|\alpha_{\hat z}\rangle$. Using \eqref{coherentstate}, we obtain
\begin{align}
|\partial_z\alpha_{\hat z}\rangle &= D(\alpha_{\hat z})\left( \epsilon a^\dagger - \frac{1}{2}\left[ \alpha_{\hat z}\eta^* + \alpha_{\hat z}^*\epsilon \right]  \right)|0\rangle, \label{derivative_coherent_1}\\
|\partial_{z^*}\alpha_{\hat z}\rangle &= D(\alpha_{\hat z}) \left( \eta a^\dagger - \frac{1}{2}\left[ \alpha_{\hat z}\epsilon^* + \alpha_{\hat z}^*\eta \right]  \right)|0\rangle. \label{derivative_coherent_2}
\end{align}
Thereby, the CSLD are
\begin{equation}
\mathcal{L}_{z} = 2D(\alpha_{\hat z})( \eta^*a + \epsilon a^\dagger  )D(\alpha_{\hat z})^\dagger, \qquad \mathcal{L}_{z^*} =\mathcal{L}_{z}^\dagger.
\end{equation}
Replacing the above relations in \eqref{QCRB_opt_cond_example}, we obtain the following optimality condition
\begin{equation}
|\eta|^2=|\epsilon|^2.
\end{equation}
To attain the QCRB, we must calculate the CSLD-QFIM. Employing \eqref{CSQFIM}, \eqref{CFM1} and \eqref{CFM2}, we obtain
\begin{equation}
\mathcal{J}_{\hat{z}}^S
=
2\left[\begin{array}{cc}
|\epsilon|^2+|\eta|^2 & 2\epsilon^*\eta \\
2\epsilon\eta^* & |\epsilon|^2+|\eta|^2
\end{array}\right].
\end{equation}
Therefore, the symmetric QCRI reads
\begin{equation}
\Cov(\hat{z}) \geq (\mathcal{J}_{\hat{z}}^S)^{-1} = \frac{1}{2(|\epsilon|^2-|\eta|^2)^2}\left[\begin{array}{cc}
|\epsilon|^2+|\eta|^2 & -2\epsilon\eta^* \\
-2\epsilon^*\eta & |\epsilon|^2+|\eta|^2
\end{array}\right].
\end{equation}
Let us note that $\mathcal{J}_{\hat{z}}^S $ is invertible, and then the symmetric CRB exists if its determinant does not vanish, that is,
\begin{equation}
\det(\mathcal{J}_{\hat{z}}^S) = 2(|\epsilon|^2 - |\eta|^2)^2 \neq 0.
\end{equation}
We arrive thus at a contradiction, since in the case where we achieve optimality, the FIM cannot be inverted. Therefore, we conclude that no measurement scheme achieves the CRB. This result is not surprising, as obtaining $z$ is equivalent to estimating its real and imaginary parts through quadrature operators $X=(a+a^\dagger)/2$ and $Y=-i(a-a^\dagger)/2$, which cannot be measured simultaneously. Let us consider that $\epsilon=1$ and $\eta=0$, and a complex statistic $\hat{t}$ corresponding to a independent homodyne measurement into the quadrature operators $X$ and $Y$ with estimator of $z$ as $\ev{X}+i\ev{Y}$, we have the following complex covariance matrix
\begin{align}
    \Cov(\hat{z})= \left[\begin{array}{cc}
    1 & 0 \\ 0 & 1
    \end{array}\right],
\end{align}
while the CSLD-QFIM is
\begin{align}
    \mathcal{J}_{\hat{z}}^S= \frac{1}{2}\left[\begin{array}{cc}
    1 & 0 \\ 0 & 1
    \end{array}\right],
\end{align}
where we can clearly see that the QCRB is not attained and that $\Cov(\hat{z})=2\mathcal{J}_{\hat{z}}^S$. An alternative is to study the problem using the CRLD-QCRI. Employing \eqref{KRCFIM}, \eqref{RFIM1_pure} and \eqref{RFIM2_pure}, and the derivatives \eqref{derivative_coherent_1} and \eqref{derivative_coherent_2}, we have obtain that
\begin{equation}
\mathcal{K}_{\hat{z}} = -2i\left[\begin{array}{cc}
|\epsilon|^2-|\eta|^2 & 0 \\
0 & -|\epsilon|^2+|\eta|^2
\end{array}\right].
\end{equation} 
Using \eqref{ComplexPureQFIM}, the CRLD-QCRI is given by
\begin{equation}
\Cov_{\hat z} \geq (\mathcal{J}_{\hat{z}}^R)^{-1} = \frac{1}{(|\epsilon|^2-|\eta|^2)^2}\left[\begin{array}{cc}
|\epsilon|^2 & -\epsilon\eta^* \\ -\epsilon^*\eta & |\eta|^2
\end{array}\right].
\end{equation}
We observe that the CSLD- and CRLD-CRIs yield different bounds on the covariance matrix, providing different information about the problem. In particular, the matrix $(\mathcal{J}_{\hat{z}}^R)^{-1}$ has rank one. This can be seen by considering $\epsilon=1$ and $\eta=0$, where 
\begin{equation}
\Cov_{\hat z} \geq (\mathcal{J}_{\hat{z}}^R)^{-1} = \left[\begin{array}{cc}
1 & 0 \\ 0 & 0
\end{array}\right].
\end{equation}
Thus, the bounds for the variance of  $z$ and $z^*$ are given by $\Cov(z) \geq 1 $ and $\Cov(z^*) \geq 0$, respectively. While the bound for $z^*$ is trivial, the bound for $z$ provides a fundamental limit for estimating $z$ and is achieved by the statistic $\hat{t}$ of measuring $X$ and $Y$ independently. This indicates that, even though the QCRB for the extended variable $\hat{z}$ cannot be reached, it is attainable when we focus on the variable $z$.

The problem can also be studied in terms of the weighted mean square error (WMSE) \eqref{CMSE}. Considering a diagonal block weighting matrix $\mathcal{W}_{\hat{z}}=diag(\mathtt{W}_z,\mathtt{W}_z^*)$, the CSLD-QFIM establishes the lower bound \eqref{SBtoCMSE}, that is, 
\begin{equation}
w^S_{\hat z} = \frac{|\epsilon|^2+|\eta|^2}{2(|\epsilon|^2-|\eta|^2)^2}(\mathtt{W}_z+\mathtt{W}_z^*),
\end{equation}
while the CRLD-QFIM defines the lower bound \eqref{RBtoCMSE} given by
\begin{equation}
w^R_{\hat z} = \frac{|\epsilon|^2+|\eta|^2}{2(|\epsilon|^2-|\eta|^2)^2}(\mathtt{W}_z+\mathtt{W}_z^*) + \frac{|\mathtt{W}_z|}{\left||\epsilon|^2-|\eta|^2 \right| }. \label{eq:QMSE_R_example}
\end{equation}
We can see that symmetric and right bounds are related by $w^R_{\hat z} =w^S_{\hat z}+|\mathtt{W}_z|/\left||\epsilon|^2-|\eta|^2 \right|$. Therefore, for a diagonal weighting error, estimation theory based on the CRLD provides the supremum, that is, the largest of the lower bounds, corresponding to the fundamental bound for the problem. The WMSE can be used as a metric to identify the optimal encoding strategy. For that, we have to set an energy upper bound $|\epsilon|^2+|\eta|^2\leq E$, and then minimize the WMSE \eqref{eq:QMSE_R_example} over $\epsilon$ and $\eta$. In particular, when $\mathtt{W}_z=1$ and $E=1$, the optimum is given by $(\epsilon,\eta)\in\{ (\pm1,0),(0,\pm1)  \}$.

The protocol can be improved using a two-mode coherent state $|\alpha_{\hat z}^{(1)}\rangle|\alpha_{\hat z}^{(2)}\rangle$, where the symmetric QCRI can be saturated \cite{Arnhem2019}. In this case, we consider the encoding $\alpha_{\hat z}^{(1)} = \epsilon_1 z + \eta_1 z^*$ and $\alpha_{\hat z}^{(2)} = \epsilon_2 z + \eta_2 z^*$ with the key $(\epsilon_1,\eta_1,\epsilon_2,\eta_2)$. Defining the two-mode displacement operator as $D(\alpha_{\hat z}^{(1)},\alpha_{\hat z}^{(2)})=D(\alpha_{\hat z}^{(1)})\otimes D(\alpha_{\hat z}^{(2)})$, we have that the CSLDs are
\begin{equation}
     \mathcal{L}_z = 2D(\alpha_{\hat z}^{(1)},\alpha_{\hat z}^{(2)}) ( \eta_1^*a_1 + \epsilon_1 a_1^\dagger + \eta_2^*a_2 + \epsilon_2 a_2^\dagger  )D(\alpha_{\hat z}^{(1)},\alpha_{\hat z}^{(2)})^\dagger, \qquad \mathcal{L}_{z^*} =\mathcal{L}_{z}^\dagger,
\end{equation}
and that the attainability condition is
\begin{equation}
    |\eta_1|^2+|\eta_2|^2=|\epsilon_1|^2+|\epsilon_2|^2.
\end{equation}
The CSLD-QFIM becomes 
\begin{equation}
\mathcal{J}_{\hat{z}}^S
=
2\left[\begin{array}{cc}
|\epsilon_1|^2+|\eta_1|^2 +|\epsilon_2|^2+|\eta_2|^2 
& 2\epsilon_1^*\eta_1 + 2\epsilon_2^*\eta_2  \\
2\epsilon_1\eta_1^* + 2\epsilon_2\eta_2^* 
& |\epsilon_1|^2+|\eta_1|^2 +|\epsilon_2|^2+|\eta_2|^2 
\end{array}\right],
\end{equation}
and its determinant is given by
\begin{equation}
    \det(\mathcal{J}_{\hat{z}}^S) = 2(|\epsilon_1|^2+|\eta_1|^2 +|\epsilon_2|^2+|\eta_2|^2 )^2 - 8|\epsilon_1^*\eta_1 + \epsilon_2^*\eta_2 |^2.
\end{equation}
Thus, in this case, it is possible to find keys that simultaneously satisfy the optimality condition and define an FIM whose determinant does not vanish. For instance, setting $(\epsilon_1,\eta_1,\epsilon_2,\eta_2)=(1,0,0,1)$ the CRB reads
\begin{equation}
    \Cov_{\hat z} \geq (\mathcal{J}_{\hat{z}}^S)^{-1} = \frac{1}{4}\left[\begin{array}{cc}
        1 & 0 \\ 0 & 1
\end{array}\right].
\end{equation}
This bound can be attained because we can estimate the real and imaginary parts of $z$ in a single experiment by performing a two-mode homodyne measurement in the operator
\begin{equation}
    Q = X\otimes \mathbb{I} + i\mathbb{I}\otimes Y,
\end{equation}
In simple words, we are using one mode for the real part of $z$ and the other for the imaginary part.

\section{Conclusion and Outlook}

In this article, we have formulated the theory of quantum estimation of complex statistics on the dependence of complex parameters, which is natively developed in the field of complex numbers. This formulation is based on manipulating the complex parameters through their conjugate extension instead of their representation in its real and imaginary parts, and using the Wirtinger calculus instead of the real calculus. This formulation states new versions of the main quantities of quantum estimation theory, that is, logarithmic derivatives, both symmetric (SLD) and right (RLD), Fisher information matrices (FIM), and quantum Cramér-Rao bounds (CRB). We define lower bounds for the weighted mean square error (WMSE), which is widely used in multi-parametric estimation.  Furthermore, working with the CRBs by block matrices, we state lower bounds for the studied statistic and not only for its conjugate extension or its real representation. All of the above results are particularized for the case of a pure state. An application in quantum communication was studied, which consists of the optimal estimation of a complex parameter encoded in a coherent state.

This theory is equivalent to its real counterpart since the main results, such as FIMs and Cramér-Rao inequalities, can be connected through the map $\ev{}_\C$. However, our theory is self-contained, that is, we can study a quantum estimation problem just by using it, independently of its real counterpart. This can be exploited in problems where the parameters are natively complex. The most direct scenario is the estimation of quantum states, processes, or measurements, which are represented by matrices with complex entries \cite{Mahler2013, Hou2016, Li2016, Pereira2018, Struchalin2018, 2009.04791, Zambrano2020, Zambrano2020_3B}, making the formulation over the complex numbers both convenient and fundamentally consistent with the underlying physics \cite{nielsen2010quantum}. Complex parameters also appear in quantum communication and metrology with continuous-variable quantum systems \cite{Huang2018,Arnhem2019,Nielsen2025}, where both coherent and squeezed states are labeled by complex amplitudes. Beyond these, variational quantum algorithms \cite{Qi2024,Cerezo2021}, such as variational quantum eigensolver (VQE) \cite{Peruzzo2014} or quantum neural networks \cite{Beer2020}, optimize a cost function over parameterized quantum states, and therefore naturally involve optimization over complex manifolds \cite{Brody2001,Chruciski2019,Hou2024}. Extensions of VQE have also been proposed for non-Hermitian Hamiltonians, where the eigenvalues are complex, and the optimization must account for this structure \cite{Xie2024}. 

Other results in quantum estimation theory can also be extended to the complex case with our approach, such as the Holevo-Cramér-Rao bound \cite{Holevo2011,Hayashi2005,Albarelli2019}. This is defined as the solution to an optimization problem and represents the lowest WMSE attainable with global measurement when an asymptotically large sample is used in the estimation. Optimization methods in complex numbers have been shown to outperform their real counterparts on certain tasks \cite{Smirnov2015, Adali2014, Hirose2012, Zhang2015, Utreras2019}, suggesting that an extension of the HCRB to the complex case could provide a computational advantage over the real formalism.  

\section{Acknowledgments}
A.\thinspace D. was supported by the Millennium Institute for Research in Optics (MIRO). L.P. was supported by the Government of Spain (Severo Ochoa CEX2019-000910-S, FUNQIP, and QEC4QEA PCI2025-163167), European Union (PASQuanS2.1, 101113690 and QEC4QEA, 101194322), Fundació Cellex, Fundació Mir-Puig, and Generalitat de Catalunya (CERCA program). M.M. was supported by ANID-PFCHA/DOCTORADO-NACIONAL/2019-21190958. 

\bibliography{bibtex.bib}

\end{document}